\documentstyle[aps]{revtex}
\begin{document}

\draft
\twocolumn[\hsize\textwidth\columnwidth\hsize\csname
@twocolumnfalse\endcsname

\title{A note on the formal structure of quantum constrained 
systems}
\author{\bf Carlo Rovelli}\address{Physics Department, 
University of Pittsburgh, USA} \address{rovelli@pitt.edu}
\date{\today}
\maketitle
\begin{abstract}
The space of the solutions of Dirac's quantum constraints cannot 
be constructed factoring the quantum state space by the 
``simple'' gauge transformations generated by the constraints.  
However, we show here that it can be constructed by factoring the 
state space by suitably defined ``complete'' gauge 
transformations.  These are generated by the action of the 
quantum constraints on individual components of the quantum 
state.
\end{abstract}

\vskip2pc]

\section{Introduction}

In classical mechanics, a gauge invariant state can be seen as an 
equivalence class of gauge noninvariant states.  Two 
gauge noninvariant states are equivalent if there is a gauge 
transformation sending one into the other.  The same fails to be 
true in quantum mechanics: Dirac's quantum constraints $C$ 
generate the gauge transformation $\psi \rightarrow e^{itC}\psi$ 
on the quantum states, but physical states cannot be seen as 
equivalence classes under the equivalence relation
\begin{equation}
\psi \sim e^{itC}\psi. 
	  \label{simple}
\end{equation}
Rather, physical states are the states which are 
annihilated by the Dirac constraints \cite{dirac}.

We show in this brief note that one {\rm can\/} see 
physical states as equivalence classes of gauge noninvariant 
states in the quantum theory as well, but under an equivalence 
relation more complicated than (\ref{simple}).  We call this 
alternative equivalence relation a ``complete'' quantum gauge 
transformation.  Roughly, a complete quantum gauge transformation 
is defined as follows.  $\phi\sim\psi$ iff there are states 
$\rho_{i}$ and $t_{i}$'s such that
\begin{eqnarray}
	\psi & = & \sum_{i}\rho_{i} \nonumber  \\
	\phi & = & \sum_{i} e^{it_{i}C} \rho_{i}. 
	\label{uno}
\end{eqnarray}
That is, a complete gauge transformation is obtained by 
independently gauge transforming linear components.  We show below 
that the space of the solutions of Dirac's constraint is 
naturally identified with the space of the equivalence classes 
defined by this equivalence relation.

In general, in an infinite dimensional Hilbert space, the Dirac 
physical states can be generalized vectors that do not belong to the 
Hilbert space.  In this case, the $\rho_{i}$ in (\ref{uno}) may 
be generalized vectors as well.  However, we shall show below 
that complete gauge equivalence can be defined in terms of the 
finite gauge operator $e^{itC}$ also without recurring to 
generalized vectors.  This fact allows us to construct the space 
of the physical states as a space of equivalence classes, 
without the need of extending the Hilbert space or using 
generalized vectors.

\section{Quantum gauge transformations in a finite dimensional
Hilbert space}

Let us assume that we have a unitary representation $U$ of a 
group $G$ in a Hilbert space $\cal H$.  In this section we 
disregard all complications due to the infinite dimensionality of 
$\cal H$.  The generators of the representation are the Dirac 
constraints, and the space of physical states ${\cal H}_{Ph}$ is 
defined as the kernel of the Dirac constraints \cite{dirac}, 
namely as the trivial representation of $G$ in $\cal H$.  States 
in ${\cal H}_{Ph}$ are gauge invariant, and represent physical 
states.  A gauge noninvariant state can roughly be seen as a 
state in a particular gauge.  Physical predictions of a classical 
gauge theory are given by gauge invariant quantities; but in 
concrete calculations, we usually employ a gauge-non-invariant 
description -- leaving the task of extracting the physical 
quantities at the end.  It would be nice to be able to do the 
same in the quantum theory, namely to work on $\cal H$ without 
recurring to ${\cal H}_{Ph}$, keeping track of gauge equivalence.

The orthogonal projection $\pi$ on ${\cal H}_{Ph}$ provides 
the natural definition of quantum gauge equivalence in $\cal H$: 
$\phi\sim\psi$ iff
\begin{equation}
	\pi(\phi) = \pi(\psi).
\label{equivalence}
\end{equation}
We have ${\cal H}_{Ph} = \frac{\cal H}{\sim}$.

What is the precise relation between this equivalence and the 
transformations generated by $U$ in $\cal H$?  Can we 
interpret this equivalence as the possibility of being gauge 
transformed, as we do for the quantum theory?  More precisely,
can we construct the equivalence relation $\sim$ directly from
$U$ without having to solve for the invariant states first? 
Clearly if there exists a $g\in G$ such that 
\begin{equation}
	\phi=U[g]\psi
\end{equation} 
then $\phi\sim\psi$.  However, the converse is not true in 
general.  Namely $\phi$ and $\psi$ can be gauge equivalent even 
if there is no $U[g]$ that maps one into the other.  

To get some intuition on how this may come about consider the 
following simple example.  Let the group $U(1)$ act on $R^{3}$ by 
generating rotations around the $z$ axis.  The invariant subspace 
is the one dimensional $z$ axis, while the space of the orbits is 
the two dimensional space of the circles parallel to the $z=0$ 
plane and centered on the $z$ axis.

Clearly, it is the linear structure of quantum mechanics that 
differentiates gauge equivalence from the fact of belonging to 
the same orbit: two orbits on the same $z=constant$ plan are in 
the same gauge equivalence class.

This example suggests that two quantum states are quantum gauge 
equivalent not only if they can be transformed into each other by 
a finite gauge transformation, but also if they can be decomposed 
into a linear combination of vectors which can be independently 
gauge transformed into each other.  We make this intuition 
concrete as follows.  We now show that $\psi$ and $\phi$ are 
equivalent (that is (\ref{equivalence}) holds), iff there exist 
vectors $\rho_{i}\in {\cal H}$ and elements $g_{i}\in G$ such 
that
\begin{eqnarray}
	\psi & = &  \sum_{i} \rho_{i},  \nonumber  \\
	\phi & = &  \sum_{i} U[g_{i}] \rho_{i}.
	\label{main}
\end{eqnarray}

To prove that (\ref{main}) implies (\ref{equivalence}) is 
immediate: it suffices to notice that $\pi U[g] = \pi$ by 
definition.  To prove the converse, we begin by proving that 
any vector $\rho$ in the kernel $K$ of $\pi$ can be written 
as 
\begin{equation}
   \rho = \sum_{i} (U[g_{i}] \rho_{i}-\rho_{i}). 
  \label{L}
\end{equation}
Let $L$ be the space of all vectors that can be written as in 
(\ref{L}).  $L$ is a linear subspace, it is let invariant 
by $U[g]$ and is contained in $K$.  Let $S$ be the subspace of 
$K$ orthogonal to $L$.  A vector $\rho$ in $S$ cannot be $U$ 
invariant because it is in $K$, therefore $\chi=U[g]\rho-\rho$ is 
different from zero.  But $S$ must be left invariant as well, 
therefore $\chi\in S$ and not in $L$, but $\chi$ is also in $L$, 
by definition of $L$, therefore $S$ is empty, $K=L$ and all 
vectors in $K$ can be written as in (\ref{L}).  

Now, if $\pi(\phi)=\pi(\psi)$, then $(\phi-\psi)\in K$, therefore 
there are $g_{i}$ and $\rho_{i}$ such that
\begin{equation}
	\phi-\psi=\sum_{i} (U[g_{i}] \rho_{i}-\rho_{i}).
\label{diff}
\end{equation}
It follows that 
\begin{equation}
	\phi-\sum_{i}U[g_{i}] \rho_{i}=\psi-\sum_{i}\rho_{i}\equiv\rho.
\label{rho}
\end{equation}
By adding $\rho$ to both sums (with a corresponding $g$ 
=identity), we have (\ref{main}), {\em QED}.

Thus, if we {\em define\/} the following equivalence relation: 
$\phi\sim\psi$ iff there exist $\rho_{i}\in{\cal H}$ and 
$g_{i}\in G$ such that (\ref{main}) holds -- then we have
	\begin{equation}
     {\cal H}_{Ph}= \frac{\cal H}{\sim}. 
\end{equation}
Intuitively, a quantum state is a linear quantum superposition of 
classical configurations (a wave function over configuration 
space).  It is therefore reasonable that we may gauge transform 
each individual component of the superposition independently, 
without changing the gauge invariant quantum state.

We call the transformation $\psi \rightarrow U[g]\psi$ a ``simple'' 
quantum gauge transformation, and the transformation
\begin{equation}
	\psi=\sum_{i}\rho_{i} \rightarrow \phi=\sum_{i}U[g_{i}] \rho_{i}
\end{equation}
a ``complete'' quantum gauge transformation.  We have proven that
physical quantum states are not equivalence classes under simple 
quantum gauge transformations, but they are equivalence classes 
under complete quantum gauge transformations.

\section{Infinite dimensional issues}

In infinite dimensional spaces, the well known infinitary 
subtleties of quantum mechanics appear.  In general, zero can be 
in the continuum spectrum of the Dirac constraints and therefore 
physical states appear as generalized states.  We have then to 
use continuum-spectrum techniques, such as Gelfand triples 
\cite{Gelfand}, or similar.  In particular, ${\cal H}_{Ph}$ is 
not a linear subspace of $\cal H$, but a linear subspace of a 
suitable closure $\overline{\cal H}$ of $\cal H$, which can be 
defined as the dual of a suitable dense subspace of $\cal H$.

In this case, the analysis of the previous section can be repeated 
with minor modifications, using $\overline{\cal H}$. $U$ acts on 
$\overline{\cal H}$ by duality.  ${\cal H}_{Ph}$ is the 
$U$-invariant subspace of $\overline{\cal H}$.  Let $L$ be the 
subspace of $\overline{\cal H}$ formed by the vectors that can be 
written in the form (\ref{L}), where now the sum may contain an 
infinite numbers of terms, and the required convergence is in 
$\overline{\cal H}$, not in $\cal H$.  Consider $S = 
\frac{\overline{\cal H}}{{\cal H}_{Ph}\oplus L}$.  As before, it 
is easy to see that $S$ is linear and $U$-invariant.  If $\rho$ 
is a non vanishing vector in $S$, it cannot be $U$ invariant 
(because it would be in ${\cal H}_{Ph}$) and $\chi=U[g]\rho-\rho$ 
is different from zero.  But $S$ is left invariant as well, 
therefore $\chi\in S$ and not in $L$, but $\chi$ is also in $L$, 
by definition of $L$, therefore $S$ is empty and $\overline{\cal 
H}={\cal H}_{Ph}\oplus L$.  

Notice that even if $\psi$ and $\phi$ are in $\cal H$, in general 
the $\rho_{i}$'s are in $\overline{\cal H}$ and not in $\cal H$.  
More precisely, the r.h.s.  of (\ref{diff}) is obviously in $\cal 
H$ if $\psi$ and $\phi$ are; but when we split the sum into the 
two sums in (\ref{rho}), the individual sums need not converge in 
$\cal H$.  Thus, $\rho$ in (\ref{rho}) may be a generalized 
vector.  Therefore we can still define ${\cal H}_{Ph}$ as the 
space of the equivalence classes of vectors in $\cal H$ under the 
equivalence relation (\ref{main}), but we must allow for 
decompositions in generalized vectors $\rho_{i}$ as well.

However, the analysis above suggests that we can avoid the 
cumbersome introduction of $\overline{\cal H}$ and generalized 
vectors altogether.  This follows from the fact that space $L$ of 
the vectors that can be written in the form (\ref{L}) {\rm is\/} 
a proper subspace of $\cal H$.  Thus, we can define $L$ first, 
and construct the linear space ${\cal H}_{Ph}$ as the space of 
the equivalence classes of vectors in $\cal H$, equivalent under 
the addition of vectors in $L$, namely as $\frac{\cal H}{L}$.  In 
other words, we can define the the equivalence relation by 
(\ref{diff}) instead than by (\ref{main}).  This is done as 
follows.

Given an infinite dimensional Hilbert space $\cal H$ and a 
unitary representation $U$ of a group $G$ over it, we define $L$ 
as the closed linear subspace of $\cal H$ formed by the vectors 
that can be written as
\begin{equation}
   \rho = \sum_{i=1}^{\infty} (U[g_{i}] \rho_{i}-\rho_{i}). 
\end{equation}
We then call two states gauge equivalent if their difference is 
in $L$, and define 
\begin{equation}
     {\cal H}_{Ph} = \frac{\cal H}{L}. 
\label{def}
\end{equation}
The space ${\cal H}_{Ph}$ is defined in this way without 
recurring to generalized vectors or other extensions of $\cal H$.  
This space is naturally isomorphic to the space of generalized 
vectors that solve the Dirac constraints.

To clarify how this may happen, consider the following.  In 
finite dimensions, if $L$ is a proper subspace of $\cal H$, then 
$L_{\perp}$ the orthogonal complement of $L$ (that is, the set of 
vectors orthogonal to $L$) is nontrivial, and 
\begin{equation}
	{\cal H} =  L_{\perp} \oplus L. 
\end{equation}
We can thus identify $L_{\perp}$ with $\frac{\cal H}{L}$.  In 
infinite dimensions, the orthogonal complement $L_{\perp}$ of a 
subspace $L$ may be trivial (contain only the zero vector) even 
if $L$ is smaller that $\cal H$.  But $\frac{\cal H}{L}$ exists 
nevertheless, and it is naturally identifiable with the space of 
{\em generalized\/} vectors perpendicular to $L$.  Gauge 
invariance of a generalized vector means being perpendicular to 
$L$.  Therefore, if we construct ${\cal H}_{Ph}$ by requiring 
gauge invariance (solving the Dirac constraints), we need 
generalized vectors. But if we if we construct ${\cal H}_{Ph}$ 
as a the space of the gauge equivalence classes, we do not need 
to introduce generalized states. 

\section{A simple example}

Let us see how this is realized in a very simple example.  Let 
$\cal H$ be the space $L_{2}[R^{2}]$ of functions $\psi(x,y)$, 
and let us have a single constraint $C = \imath 
\frac{\partial}{\partial y}$.  The group $G$ is the abelian group 
$R$ that acts on $\cal H$ by displacing functions in the $y$ 
direction: $U(a)\psi(x,y) = \psi(x,y+a)$.  What is the space $L$?  
All the vectors of the form $\rho= U(a)\psi-\psi$ satisfy
\begin{equation}
	\int dy\ \rho(x,y) = 0,
	\label{K}
\end{equation}
and any vector that satisfies (\ref{K}) can be approximated in 
norm by a sequence of vectors of the form $\rho= U(a)\psi-\psi$.  
Therefore equation (\ref{K}) defines the subspace $L$ in $\cal 
H$.  Notice that $L$ is a well defined proper subspace of $\cal 
H$.  According to our definition, two functions $\psi(x,y)$ and 
$\phi(x,y)$ are gauge equivalent if their difference is in $L$, 
namely if
\begin{equation}
	\int dy\ \left(\psi(x,y)-\phi(x,y)\right) = 0.
\end{equation}
We can characterize (a dense subspace of) the equivalence 
classes by functions of one variable
\begin{equation}
	\psi(x) \equiv \int dy\ \psi(x,y) = \int dy\ \phi(x,y).
\end{equation}
Therefore the map $\pi:{\cal H}\mapsto {\cal H}_{Ph}$ 
is integration in $dy$ and ${\cal H}_{Ph}$ is formed by square 
integrable functions of $x$ alone.%
	  \footnote{We recall that, contrary to what often 
	  stated, $\cal H$ naturally induces a scalar product 
	  in ${\cal H}_{Ph}$, although some work may be 
	  required to write this scalar product explicitly.  In 
	  the present example, for instance, a spectral 
	  decomposition theorem states that $\cal H$ can be 
	  written as a direct integral of Hilbert spaces ${\cal 
	  H}=\int dk\ {\cal H}_{k}\ $ \cite{Guichardet}.  $C$ 
	  is diagonal in each ${\cal H}_{k}$ with eigenvalue 
	  $k$.  The point here is that ${\cal H}_{0}$ is 
	  naturally identified with ${\cal H}_{Ph}$, {\em 
	  and\/} it carries a scalar product.}

A ``simple'' gauge transformation is a rigid displacement of 
$\psi(x,y)$ along the $y$ axis.  A ``complete'' gauge 
transformation is the addition of any function with vanishing 
$\int dy$ integral.  The physical space is the space of the 
functions $\psi(x,y)$ modulo these complete gauge 
transformations.

\section{Conclusions}

We have introduced the notion of complete quantum gauge 
transformation.  In the finite dimensional case, a complete gauge 
transformation is obtained by decomposing a vector in components 
and acting with the exponentiated constraints on each component 
independently.  Namely, two vectors are gauge equivalent if 
equation (\ref{main}) holds.  In the infinite dimensional case, 
the same can be done allowing decompositions on bases of 
generalized vectors as well.  Equivalently, two states $\psi$ and 
$\phi$ are quantum gauge equivalent if their difference is in the 
closure of the space of the linear combinations of the vectors 
$U(g)\psi-\psi$, namely if (\ref{diff}) holds.  Using this second 
strategy, there is no need of introducing generalized vectors.  
Dirac's physical state space ${\cal H}_{Ph}$ can be obtained as 
the space of the equivalence classes of states, under complete 
quantum gauge transformations.

In the classical hamiltonian theory of constrained systems, one 
has to take two steps in order to reduce the full phase space 
$\Gamma$ to the physical phase space $\Gamma_{Ph}$.  First, solve 
the constraint; that is, find the constraint surface $C$ in 
$\Gamma$.  Second, factor away the gauge transformation; that is, 
define $\Gamma_{Ph}$ as the space of the gauge orbits in $C$.  
Dirac showed that in the quantum theory a single step is 
sufficient: the physical states are the ones that solve the 
quantum constraints.  Here we have shown that one can take this 
single step also by factoring away (complete) quantum gauge 
transformations.  Thus, in the classical theory we find the 
physical states by solving the constraints {\em and\/} factoring 
away the gauge transformations.  In the quantum theory we find 
the physical states by solving the constraints {\em or\/} 
factoring away the gauge transformations.

The existence of this alternative strategy for dealing with quantum 
constraints is of some interest by itself, as a small further 
clarification of the structure of constrained quantum systems.  
But it might also have some practical application.  First, it 
may allow us to avoid the cumbersome use of generalized vectors 
or extensions of the Hilbert space, as illustrated in sections 
III and IV. (Techniques for explicitly constructing the scalar 
product in the physical Hilbert space are still required, 
however.)  Second, there are systems in which the Dirac operator 
is ill defined, but its exponentiated action is well defined.  A 
typical example is the diffeomorphism constraint in loop quantum 
gravity (see \cite{loops} and references therein).  In these 
cases, finite gauge transformations are very natural objects in 
the quantum theory.  Finally, when the physical state space is 
too complicated to be constructed explicitly, the above 
construction opens the possibility of working with gauge 
noninvariant states without loosing track of gauge equivalence.  
In loop quantum gravity, steps towards the construction of the 
exponentiated hamiltonian constraint operator have been taken  
\cite{surfaces}.  An orbit generated by this operator 
corresponds to the coordinate-time evolution of a quantum states of 
gravity, or to a ``quantum spacetime''.  This note may help 
clarifying its physical interpretation.  These applications 
will be discussed elsewhere.

\vskip1cm

My thanks to Jerzy Lewandowski, particularly for repeating to me 
so many times that ${\cal H}_{Diff}$ is not $\frac{\cal 
H}{Diff}$, that I finally understood; and to Roberto DePietri for 
extensive discussions and help.  Support for this work came from 
NSF grant PHY-95-15506.

\end{document}